\documentclass[aps,floats,twocolumn]{revtex4}
\usepackage{epsf, bm}
\usepackage{amsfonts}
\usepackage{amsmath}
\usepackage{amssymb}
\usepackage[latin2]{inputenc}
\usepackage[T1]{fontenc}
\usepackage{epsfig}
\usepackage{mathrsfs}
\usepackage{verbatim}

\newcommand{\m}{{\bm m}}
\newcommand{\n}{{\bm n}}
\newcommand{\s}{{\bm s}}
\renewcommand{\r}{{\bm r}}
\renewcommand{\j}{{\bm j}}
\renewcommand{\i}{{\bm i}}
\renewcommand{\k}{{\bm k}}

\newcommand{\tr}{\text{tr}}

\newcommand{\id}{\text{id}}

\newcommand{\mh}{\mathcal{H}}

\newtheorem{thm}{Theorem}
\newtheorem{df}{Definition}
\newtheorem{lem}{Proposition}

\newlength{\dinwidth}
\newlength{\dinmargin}
\DeclareMathAlphabet{\scr}{U}{rsfs}{m}{n}

\begin{document}

\title{Entanglement and quantum groups}
\author{J. K. Korbicz$^{1,2,3,4}$\footnote{jkorbicz@mif.pg.gda.pl},
J. Wehr$^{5}$, and M. Lewenstein$^{2,6}$}

\affiliation{$^1$ Dept. d'Estructura i Constituents de la Mat\`eria,
Universitat de Barcelona, 647 Diagonal, 08028 Barcelona, Spain}

\affiliation{$^2$ ICFO--Institut de Ci\`{e}ncies
Fot\`{o}niques, Mediterranean Technology Park, 08860
Castelldefels (Barcelona), Spain}

\affiliation{$^3$ Faculty of Applied Physics
and Mathematics, Technical University of Gda\'{n}sk, 80-952
Gda\'{n}sk, Poland}

\affiliation{$^4$National Quantum Information Centre of
Gda\'{n}sk, 81-824 Sopot, Poland}

\affiliation{$^5$ Department of Mathematics, University of Arizona,
617 N. Santa Rita Ave., Tucson, AZ 85721-0089, USA}

\affiliation{$^6$ICREA-- Instituci\'o Catalana  de Recerca i Estudis Avan\c cats,
E-08010 Barcelona, Spain}

\begin{abstract}
We describe quantum mechanical entanglement in terms of compact quantum groups.
We prove an analog of positivity of partial transpose (PPT) 
criterion and formulate a Horodecki-type Theorem.
\end{abstract}

\maketitle
\section{Introduction}
Quantum entanglement has been one of
the most challenging problems of modern quantum 
mechanics. We briefly recall here the definition,
referring the reader to Ref. \cite{mama} for
a complete overview. We consider a composite
quantum system, composed of two subsystems.
If the individual systems are described by 
Hilbert spaces $\mh$ and $\widetilde\mh$, then,
according to the postulates of quantum theory,
the composite system is described by the tensor
product $\mh\otimes\widetilde\mh$. In this work
we will be interested only in those cases when
$\mh$ and $\widetilde\mh$
can be chosen to be finite-dimensional, e.g.
for a pair of spins. A state of the system is
represented by a density matrix $\varrho$, \
that is a positive operator from 
$\mathcal L(\mh\otimes\widetilde\mh)$, normalized
by $\tr \varrho=1$. In an obvious way, if
$\varrho$ and $\tilde\varrho$ are states
of the individual subsystems, then the product 
$\varrho\otimes\widetilde\varrho$ is a state
of the compound system. And so are convex combinations
of such products:
\begin{equation}\label{srep}
\sum_\lambda p_\lambda\varrho_\lambda\otimes\tilde\varrho_\lambda,
\quad p_\lambda\geqslant 0.
\end{equation}
It turns out, however, that not every state of the whole
system can be represented in the above 
form \cite{schroedinger}---the state space of the composite
system is strictly larger. Those states which admit the above
representations are called {\it separable} or {\it classically
correlated} and those which do not---{\it entangled}. Intuitively,
entangled states reflect a very strong correlation between the 
subsystems. So strong that it even violates 
a certain locality principle \cite{bell}. From a more practical 
point of view, entanglement is a resource for quantum information
processing, e.g. for teleportation, computation, cryptography \cite{mama}.
However, the question of an efficient characterization of
entangled states turned out to be a very hard task: 
despite many attempts, this
problem still does not possess a satisfactory solution
\cite{mama}.
 
In previous works we proposed \cite{pra}
and developed \cite{cmp} a novel method
of studying (generalized) entanglement
using abstract harmonic analysis on
ordinary compact groups. 
The core of the method is identification of 
the Hilbert spaces describing the system
with representation spaces of some compact groups.
Then with the help of Fourier transform 
we switch from density matrices to 
continuous positive definite functions 
on direct product of the groups
and define and study entanglement in terms
of those functions.
The main result of that approach
\cite{cmp} is a Horodecki-type
Theorem \cite{horodeccy}, 
characterizing entanglement 
of positive definite functions in terms of
positive definiteness preserving maps
of continuous functions.

In this note we show how the above
classical analysis can be extended 
to compact quantum groups (CQG). 

\section{Compact Quantum Groups and
their direct products}\label{basics}
 
We begin with the notation and 
briefly recall some basic facts.
We follow the approach of Woronowicz
\cite{woron_pseudo,woron_pre,piotrek}.
Let $(A,\Delta_A)$ and $(B,\Delta_B)$
be two compact quantum groups, where
$A,B$ are unital $C^*$-algebras and $\Delta_A,\Delta_B$
are the coproducts. Let
$\{u^\alpha\}$ and $\{v^\beta\}$ be
the complete families of irreducible unitary
corepresentations of $(A,\Delta_A)$ and $(B,\Delta_B)$
respectively. 
Just like in the classical theory,
all such corepresentations are
finite-dimensional \cite{woron_pre}.
We denote by $\mh_\alpha$ and $\widetilde\mh_\beta$
the carrier Hilbert spaces of $u^\alpha$ and $v^\beta$
respectively, so that $u^\alpha$ and $v^\beta$ are unitary
elements of $\mathcal{L}(\mh_\alpha)\otimes A$ and
$\mathcal{L}(\widetilde\mh_\beta)\otimes B$ respectively.
 Fixing once and forever orthonormal bases
$\{e^\alpha_i\}_{i=1,\dots,n_\alpha}$, $n_\alpha=\text{dim}\mh_\alpha$
and $\{\tilde e^\beta_k\}_{k=1,\dots,m_\beta}$, 
$m_\beta=\text{dim}\widetilde\mh_\beta$
in each carrier space $\mh_\alpha$ and $\widetilde\mh_\beta$,
$u^\alpha$, $v^\beta$ 
can be identified with $n_\alpha\times n_\alpha$ and $m_\beta\times m_\beta$,
matrices $[u^\alpha_{ij}]$, $[v^\beta_{lk}]$ with entries in $A$  and $B$ respectively.
They satisfy comultiplication rule: 
$\Delta_Au^\alpha_{ij}=\sum_ru^\alpha_{ir}\otimes u^\alpha_{rj}$,
and analogously for $\Delta_B v^\beta_{lk}$.

Let $\mathcal A$ ($\mathcal B$) be
a linear span of all
matrix elements $u^\alpha_{ij}$ ($v^\beta_{kl}$)
of all irreducible corepresentations of $(A,\Delta_A)$
(respectively $(B,\Delta_B)$). This is an analog of the
algebra of polynomial functions on an ordinary group.
It is a dense $*$-subalgebra of $A$ ($B$), closed
with respect to the comultiplication, and carrying 
structure of a Hopf algebra \cite{woron_pre,koornwinder}.
We recall (see e.g. \cite{woron_pre})
how counit and coinverse (antipode) maps,
defined on the above Hopf algebra,
act respectively on the matrix elements:
\begin{equation}
\varepsilon_A(u^\alpha_{ij})=\delta_{ij},\quad 
\kappa_A(u^\alpha_{ij})=u^{\alpha\,*}_{ji},\label{ek}
\end{equation}
and similarly for $\varepsilon_B,\kappa_B$ defined on
$\mathcal B$. 

The main object of our study will be a direct product:
\begin{eqnarray}
& & (A,\Delta_A)\times (B,\Delta_B)=(A\otimes B, \Delta),\label{D0}\\
& & \Delta:=(\id\otimes\sigma_{AB}\otimes\id)(\Delta_A\otimes\Delta_B),
\label{D}
\end{eqnarray}
where $\sigma_{AB}\colon A\otimes B\to B\otimes A$ is the flip operator,
and the tensor products are the minimal ones.
The complete family of unitary irreducible corepresentations $\{U\}$ of
$(A\otimes B, \Delta)$ can be chosen
in the following form: 
\begin{equation}\label{prodrep}
U^{\alpha\beta}_{ikjl}:=u^\alpha_{ij}\otimes v^\beta_{kl},
\end{equation}
given by $(n_\alpha m_\beta)\times (n_\alpha m_\beta)$
matrices with entries from $A\otimes B$ (note the labeling).
Using definition (\ref{D}) we check that
they satisfy the right comultiplication rule:
\begin{equation}\label{comprod}
\Delta U^{\alpha\beta}_{ikjl}=\sum_{r,s}U^{\alpha\beta}_{ikrs}\otimes U^{\alpha\beta}_{rsjl}.
\end{equation}
The Hopf algebra, associated with the direct product
(\ref{D0})-(\ref{D}) is given by the algebraic tensor product
$\mathcal A\otimes_{\text{alg}}\mathcal B$. The counit and coinverse are
naturally defined on it by 
\begin{equation}
\varepsilon:=\varepsilon_A\otimes\varepsilon_B,\quad
\kappa:=\kappa_A\otimes \kappa_B.\label{ek2}
\end{equation}

\section{Quantum Fourier transforms of density matrices}
Let us consider density matrices $\varrho$ on
$\mh_\alpha\otimes \widetilde\mh_\beta$, i.e.
$\varrho\in\mathcal{L}(\mh_\alpha\otimes\widetilde\mh_\beta)$,
$\varrho\geqslant0$, $\tr\varrho=1$. We perform the following
transform (cf. Refs. \cite{pra,cmp} and see \cite{u1}):
\begin{equation}\label{qft}
\varrho\mapsto\hat\varrho:=\sum_{i,\dots,l}\varrho_{ikjl}U^{\alpha\beta}_{jlik}
=(\tr\otimes\id)\varrho U^{\alpha\beta}
\end{equation}
which associates with $\varrho$ an element of $\mathcal A\otimes_{\text{alg}}\mathcal B$.
The indices $i,j$ refer here to the Hilbert space $\mathcal H_\alpha$, while
$k,l$ to $\widetilde\mh_\beta$ (cf. definition (\ref{prodrep})).
Before describing entanglement,
we show how positivity
and normalization of $\varrho$ are encoded in
$\hat\varrho$.  

Since, by definition (cf. Eqs.~(\ref{ek}), (\ref{ek2})),
$\varepsilon(U^{\alpha\beta}_{ikjl})=\varepsilon_A(u^\alpha_{ij})
\varepsilon_B(v^\beta_{kl})=\delta_{ij}\delta_{kl}$ we obtain that:
\begin{equation}\label{norm}
\varepsilon(\hat\varrho)=\tr\varrho=1.
\end{equation}

To describe the positivity property, 
let $h=h_A\otimes h_B$ be a unique Haar measure on the product
$(A\otimes B, \Delta)$ \cite{woron_pseudo}. For convenience
we define the following functionals on
$A\otimes B$: $ah(b):=h(ba)$, $ha(b):=h(ab)$, $a,b\in A\otimes B$.
Then $\hat\varrho$ satisfies an analog of positive
definiteness:
\begin{equation}\label{posdef}
\big(a^*h\kappa\otimes ha\big)\Delta\hat\varrho\geqslant 0\quad
\text{for any}\ \ a\in A\otimes B.
\end{equation}
Note that from definitions (\ref{ek}), (\ref{prodrep}), and (\ref{ek2})),
$\kappa(U^{\alpha\beta}_{ikjl})=U^{\alpha\beta\, *}_{jlik}$.

To prove statement (\ref{posdef}), we first use Eq. (\ref{comprod}) and then the above 
quoted property of the coinverse:
\begin{eqnarray}
& &\big(a^*h\kappa\otimes ha\big)\Delta\hat\varrho=\nonumber\\
& &=\sum_{i,\dots,l}\varrho_{ikjl}\sum_{r,s} h\big(\kappa(U^{\alpha\beta}_{jlrs})a^*\big)
h\big(aU^{\alpha\beta}_{rsik}\big)\nonumber\\
& &= \sum_{r,s}\sum_{i,\dots,l}\varrho_{ikjl}h\big(U^{\alpha\beta\, *}_{rsjl}a^*\big)
h\big(aU^{\alpha\beta}_{rsik}\big)\nonumber\\
& &=\sum_{r,s}\sum_{i,\dots,l}\varrho_{ikjl}\overline{h\big(aU^{\alpha\beta}_{rsjl}\big)}
h\big(aU^{\alpha\beta}_{rsik}\big)\nonumber\\
& &\equiv \sum_{r,s}\langle \psi_{rs}|\varrho \psi_{rs}\rangle\geqslant 0,
\end{eqnarray}
where $\psi_{rs}:=\sum_{i,k}\overline{h\big(aU^{\alpha\beta}_{rsik}\big)}\,
e^\alpha_i\otimes \tilde e^\beta_k\in\mh_\alpha\otimes\widetilde\mh_\beta$,
with the overbar denoting complex conjugate.
We used the identity $h(a^*)=\overline{h(a)}$ and by $\langle\cdot |\cdot \rangle$ we denote
the standard scalar product in the corresponding Hilbert spaces.

The main weakness of the proposed definition of 
positive definiteness
(\ref{posdef}) is that it can be formulated only on the respective Hopf
algebras, since generally coinverse cannot be prolonged
to the whole of the quantum group. 
One possible way out is to take norm-closure 
in $A\otimes B$ of the set of positive definite elements.
However, for the purpose of this note we will not consider
such a closure and continue with purely algebraic considerations.

Consider a separable $\varrho\in\mathcal{L}(\mh_\alpha\otimes\widetilde\mh_\beta)$,
i.e. a density matrix $\varrho$ representable
as the following finite convex combination:
\begin{equation}\label{rosep}
\varrho=\sum_\lambda p_\lambda \varrho^\alpha_\lambda\otimes\varrho^\beta_\lambda,
\quad\varrho^\alpha_\lambda, \varrho^\beta_\lambda\geqslant 0.
\end{equation}
Then its transform $\hat\varrho$ is
also {\it separable}, that is \cite{u2}:
\begin{equation}\label{sep}
\widehat\varrho=\sum_\lambda p_\lambda \hat\varrho^\alpha_\lambda
\otimes\hat\varrho^\beta_\lambda,
\end{equation}
where for all $\lambda$, $\hat\varrho^\alpha_\lambda\in\mathcal A$
and $\hat\varrho^\beta_\lambda\in\mathcal B$
satisfy the normalization (\ref{norm}) and
the positive definiteness (\ref{posdef})
conditions on $(A,\Delta_A)$ and $(B,\Delta_B)$
respectively.

We would like to have the converse of the above fact.
To that end we construct an inverse
(in some sense to be clarified later)
of the transformation (\ref{qft}).
Let $F^\alpha$, respectively $\widetilde F^\beta$,
intertwine the second contragredient representation 
$u^{\alpha cc}:=(\id\otimes\kappa_A^2)u^\alpha$
with $u^\alpha$, respectively
$v^{\beta cc}$
with $v^\beta$ \cite{woron_pseudo,koornwinder}:
\begin{eqnarray}\label{k2}
& &(\id\otimes\kappa_A^2)u^\alpha=\big(F^\alpha\otimes\id\big) 
u^\alpha \big((F^\alpha)^{-1}\otimes\id\big),\\
& &(\id\otimes\kappa_B^2)v^\beta=\big(\widetilde F^\beta\otimes\id\big) 
v^\beta \big((\widetilde F^\beta)^{-1}\otimes\id\big).\label{k2'}
\end{eqnarray}
Operators $F^\alpha$, $\widetilde F^\beta$ are invertible, positive and
uniquely fixed by the condition:
$\tr F^\alpha=\tr(F^\alpha)^{-1}>0$,
and analogously for $\widetilde F^\beta$
\cite{woron_pre, koornwinder}. Then
$F^{\alpha\beta}:=F^\alpha\otimes \widetilde F^\beta>0$
intertwines $(\id\otimes\kappa^2)U^{\alpha\beta}$ with
$U^{\alpha\beta}$. Now, given an arbitrary $a\in A\otimes B$ we define
for each $\alpha,\beta$ an operator $\hat a(\alpha\beta)$
on $\mh_\alpha\otimes\widetilde\mh_\beta$ by \cite{woron_pre,piotrek}:
\begin{eqnarray}
& & \hat a(\alpha\beta):=(\id\otimes ah)
\frac{1}{\sqrt{F^{\alpha\beta}}}U^{\alpha\beta\;\dagger}\sqrt{F^{\alpha\beta}},\label{qft2}\\
& & \hat a(\alpha\beta)_{ikjl}=\sum_{m,\dots,s}
(F^{\alpha\beta})^{-\frac{1}{2}}_{ikmn}h\Big(U^{\alpha\beta\, *}_{rsmn} \,a\Big)
(F^{\alpha\beta})^{\frac{1}{2}}_{rsjl}.\nonumber
\end{eqnarray}

Since $U^{\alpha\beta}$ is irreducible,
transformation (\ref{qft2}) is onto.
Recall that matrix element of irreps satisfy
deformed orthonormality relations \cite{woron_pseudo, koornwinder}:
\begin{eqnarray}
& &h_A\big(u^{\alpha\, *}_{ij}u^{\alpha'}_{i'j'}\big)
=\frac{\delta^{\alpha\alpha'}}{\tr F_\alpha}(F^\alpha)^{-1}_{i'i}\delta_{jj'}\\
& &h_A\big(u^\alpha_{ij}u^{\alpha'\, *}_{i'j'}\big)
=\frac{\delta^{\alpha\alpha'}}{\tr F_\alpha}\delta_{ii'}F^\alpha_{j'j}\, ,
\end{eqnarray}
and analogously for the irreps of $(B,\Delta_B)$.
Thus:
\begin{equation}\label{dist}
\varrho=(\tr F^{\alpha\beta})\;
\sqrt{F^{\alpha\beta}}\,\hat{\hat\varrho}(\alpha\beta)\,\sqrt{F^{\alpha\beta}}
\end{equation}
and one can recover $\varrho$ from its transform $\hat\varrho$.

Now let $a$ belong to $\mathcal A\otimes_{\text{alg}}\mathcal B$.
We show that if $a$ is {\it positive definite}, i.e. satisfies
(cf. Eq. (\ref{posdef}))
\begin{equation}\label{posdef2}
\big(b^*h\kappa\otimes hb\big)\Delta a\geqslant 0
\end{equation}
for any $b$ from $A\otimes B$, then
$\hat a(\alpha\beta)\geqslant 0$
for all $\alpha, \beta$. To better understand the condition
(\ref{posdef2}), recall that for an ordinary compact group $G$, when
the relevant $C^*$-algebra is just the standard 
$C^*$-algebra of continuous functions on $G$, 
$h(f)=\int dg f(g)$, $dg$ being the Haar measure on $G$,
$(\kappa f)(g)=f(g^{-1})$, $f^*(g)=\overline{f(g)}$,
and $(\Delta f)(g,h)=f(gh)$, Eq.~(\ref{posdef2}) 
is just the standard positive definiteness condition:
$\int\int dg dh\, \overline{b(g)}b(h)a(g^{-1}h)\geqslant 0$
(see Refs.~\cite{pra,cmp} for the Fourier analysis of density matrices and 
separability on ordinary compact groups).

For more transparency, let us
introduce compound indices
${\bm i}\equiv (ik)$ pertaining
to $\mh_\alpha\otimes\widetilde\mh_\beta$.
Since $h$ is normalized, $h(I)=1$, and
$h(\kappa(a))=a$, we can rewrite
$h\Big(U^{\alpha\beta\, *}_{{\bm j}{\bm i}} \,a\Big)\equiv
h\Big(U^{\alpha\beta\, *}_{jlik} \,a\Big)$
as:
\begin{eqnarray}
& &h\Big(U^{\alpha\beta\, *}_{\j\i} a\Big)
h(\kappa(I))=h\kappa\Big[h\Big(U^{\alpha\beta\, *}_{\j\i} a\Big)I\Big]\nonumber\\
& &=\big(h\kappa\otimes h)\Delta\Big(U^{\alpha\beta\, *}_{\j\i} a\Big)
=\big(h\kappa\otimes h)\Delta U^{\alpha\beta\, *}_{\j\i}\Delta a\nonumber\\
& &=\sum_{\bm r}
\Big[\kappa^2(U^{\alpha\beta}_{{\bm r}\j})h\kappa\otimes h U^{\alpha\beta\, *}_{{\bm r}\i}\Big]
\Delta a\nonumber\\
& &= \sum_{\r,\m,\n}
F^{\alpha\beta}_{{\bm r}{\bm m}}
(F^{\alpha\beta})^{-1}_{{\bm n}{\bm j}}
\Big[U^{\alpha\beta}_{{\bm m}{\bm n}} h\kappa\otimes h U^{\alpha\beta\, *}_{{\bm r}\i}\Big]\Delta a.
\label{last}
\end{eqnarray}
where in the second line we used the invariance of the Haar measure:
$\big(h\otimes\id)\Delta a=h(a)I=\big(\id\otimes h)\Delta a$, and then
Eq. (\ref{comprod}) and the identities: $\kappa(ab)=\kappa(b)\kappa(a)$,
$U^{\alpha\beta\, *}_{jlik}=\kappa(U^{\alpha\beta}_{ikjl})$, and finally
Eqs. (\ref{k2})-(\ref{k2'}). Substituting Eq.~(\ref{last}) into the definition
(\ref{qft2}) we obtain:
\begin{eqnarray}
& &\hat a(\alpha\beta)_{\i\j}=\sum (F^{\alpha\beta})^{-\frac{1}{2}}_{\i\m}
(F^{\alpha\beta})^{\frac{1}{2}}_{\r\j}(F^{\alpha\beta})^{-1}_{\n\r}
F^{\alpha\beta}_{{\bm r'}{\bm m'}}\times\nonumber\\
& &\times\Big[U^{\alpha\beta}_{{\bm m'}{\bm n}} h\kappa\otimes h U^{\alpha\beta\, *}_{{\bm r'}\m}\Big]\Delta a\nonumber\\
& &=\sum\bigg[(F^{\alpha\beta})^{\frac{1}{2}}_{{\bm s}{\bm m'}}U^{\alpha\beta}_{{\bm m'}{\bm n}} (F^{\alpha\beta})^{-\frac{1}{2}}_{\n\j}h\kappa\otimes\nonumber\\
& &\otimes h
(F^{\alpha\beta})^{\frac{1}{2}}_{{\bm r'}{\bm s}}U^{\alpha\beta\, *}_{{\bm r'}\m}(F^{\alpha\beta})^{-\frac{1}{2}}_{\i\m}\bigg]\Delta a\nonumber\\
& &\equiv\sum_{{\bm s}}\Big[U^{F}_{{\bm s}\j}h\kappa\otimes
h U^{F\, *}_{{\bm s}\i}\Big]\Delta a,
\end{eqnarray}
where $U^F:=\sqrt{F^{\alpha\beta}}U^{\alpha\beta}\frac{1}{\sqrt{F^{\alpha\beta}}}$
and we used hermiticity of  $F^{\alpha\beta}$.
Hence, for an arbitrary vector $\psi\in\mh_\alpha\otimes\widetilde\mh_\beta$:
\begin{eqnarray}
& &\langle \psi |\hat a(\alpha\beta) \psi\rangle=\nonumber\\
& & =\sum_{{\bm s}}
\bigg[\sum_{\j}\psi_{\j}U^F_{{\bm s}\j}\:h\kappa\otimes
h \sum_{\i}\overline \psi_{\i}U^{F\, *}_{{\bm s}\i}\bigg]\Delta a\nonumber\\
& &\equiv\sum_{\s}\big(b^*_{\s}h\kappa\otimes hb_{\s})\Delta a\geqslant 0\label{pos}
\end{eqnarray}
by positive definiteness of $a$ (cf. Eq.(\ref{posdef2})).

Combining the above facts, given by Eqs. (\ref{posdef}), (\ref{dist}), and (\ref{pos}), we
obtain the following:

\begin{lem}\label{p1}
An operator $\varrho$ acting in the carrier space
$\mh_\alpha\otimes\widetilde\mh_\beta$ is positive
if and only if its transform $\hat\varrho$ is
positive definite, i.e. satisfies condition (\ref{posdef}).
\end{lem}

\noindent The $\hat{\phantom a}$-dual of the above Proposition also holds:

\begin{lem}\label{p2}
An element $a$ of the associated Hopf algebra 
$\mathcal A\otimes_{\text{alg}}\mathcal B$ is positive 
definite if and only if 
$\hat a(\alpha\beta)\geqslant 0$
for every irrep $\alpha, \beta$.
\end{lem}

The proof in one direction readily follows from
Eq. (\ref{pos}).  
To prove in the other, observe that, by definition, 
$a\in\mathcal A\otimes_{\text{alg}}\mathcal B$ is a finite
linear combination of matrix elements $U^{\alpha\beta}_{ikjl}$:
$a=\sum_{\alpha,\beta}\sum_{\i,\j} a^{\alpha\beta}_{\i\j}U^{\alpha\beta}_{\j\i}$.
Then:
\begin{equation}
\big(b^*h\kappa\otimes hb\big)\Delta a=\sum_{\alpha\beta}\sum_{\i,\j,\k}
a^{\alpha\beta}_{\i\j}h\big(b U^{\alpha\beta}_{\k\i}\big)
\overline{h\big(b U^{\alpha\beta}_{\k\j}\big)}.
\end{equation}
If all the matrices $a^{\alpha\beta}_{\i\j}$ are positive definite,
then the above sums are positive and hence 
$a$ is positive definite. But by Eq. (\ref{dist})
$a^{\alpha\beta}=(\tr F^{\alpha\beta})
\sqrt{F^{\alpha\beta}}\,\hat a(\alpha\beta)\,\sqrt{F^{\alpha\beta}}$.
$\Box$

\section{Separability}
In an obvious way the notion of separability
given by Eq. (\ref{sep}) applies to an arbitrary
element of $\mathcal A\otimes_{\text{alg}}\mathcal B$.
The same remark applies here as to the positive definiteness:
Since coinverse
$\kappa$ has generally no extension to the whole
algebra $A\otimes B$, a way of extending
separability to the whole group would be to
consider a norm-closure in $A\otimes B$ 
of the set of separable states. Again, we will
not pursue this line here and will be satisfied with
purely algebraic facts.

We are ready to prove the following fact, justifying
the use of compact quantum groups in the study of entanglement
(see Refs. \cite{pra,cmp} for a
classical analog):

\begin{lem} \label{chuj} A density matrix
$\varrho\in\mathcal{L}(\mh_\alpha\otimes\widetilde\mh_\beta)$
is separable if and only if its transform
$\hat\varrho$ is separable in $\mathcal A\otimes_{\text{alg}}\mathcal B$.
\end{lem}

The implication in
one direction we have already shown (cf. Eq.~(\ref{sep})).
Now assume that $\hat\varrho$ is separable:
\begin{equation}
\hat\varrho=\sum_\lambda p_\lambda a_\lambda\otimes b_\lambda,
\end{equation}
where $a_\lambda,b_\lambda$ are positive definite for each $\lambda$.
Then $\hat{\hat\varrho}$ is separable too,
which immediately follows from Eqs.
(\ref{prodrep}), (\ref{qft2}), and Proposition \ref{p1}
applied to $a_\lambda,b_\lambda$:
\begin{equation}
\hat{\hat\varrho}(\alpha\beta)=\sum_\lambda p_\lambda \hat a_\lambda(\alpha)\otimes
\hat b_\lambda(\beta),
\quad\hat a_\lambda(\alpha),\hat b_\lambda(\beta)\geqslant 0.
\end{equation}
By Eq. (\ref{dist}):
\begin{equation}
\varrho=\big(\tr F^\alpha\tr \widetilde F^\beta\big)
\sqrt{F^{\alpha}}\otimes \sqrt{\widetilde F^{\beta}}
\hat{\hat\varrho}
\sqrt{F^{\alpha}}\otimes \sqrt{\widetilde F^{\beta}},
\end{equation}
so $\varrho$ is separable.$\Box$

Proposition 3 is a nice theoretical separability criterion, that allows to
conclude about separability properties of $\varrho$ is the separability 
properties  of $\hat\varrho$ are known, and vice versa. So far, we have 
been in fact mainly carrying over the results know for quantum mechanical 
states to their transforms, either on compact (standard \cite{cmp} or quantum) 
groups. Still, having an explicit separable form of $\hat\varrho$ 
implies immediately separability of $\varrho$.

Using the same technique, combined with the fact that
$\mathcal A\otimes_{\text{alg}}\mathcal B\ni 
a=\sum^{finite}_{\alpha,\beta}\sum_{\i,\j} a^{\alpha\beta}_{\i\j}
U^{\alpha\beta}_{\j\i}$, where $a^{\alpha\beta}=(\tr F^{\alpha\beta})
\sqrt{F^{\alpha\beta}}\,\hat a(\alpha\beta)\,\sqrt{F^{\alpha\beta}}$,
we show the following basic:

\begin{lem}\label{gen}
An element $a\in\mathcal A\otimes_{\text{alg}}\mathcal B$
is separable if and only if the operators $\hat a(\alpha\beta)$ are
separable for every irrep $\alpha,\beta$.
\end{lem}

By Propositions \ref{p1} and \ref{gen}, positive definite (separable) 
elements of the associated Hopf algebra
generate a family of positive (separable) operators,
acting in the carrier spaces of irreps of the product $(A\otimes B, \Delta)$.
Moreover, each (separable) density matrix can be obtained this way
(cf. Eq. (\ref{dist})). Thus, in analogy with the classical case \cite{pra,cmp},
a description of separable elements at the level of quantum group
would provide a description of separable states in all dimensions,
where the given quantum group has irreducible corepresentations.
Motivated by this observation we state:

\begin{df}[CQG Separability Problem]
Given a positive definite element of
the Hopf algebra
$\mathcal A\otimes_{\text{alg}}\mathcal B$,
associated with the group $(A\otimes B, \Delta)$,
decide whether it is separable or not.
\end{df}

Now we derive an analog of positivity
of partial transpose criterion (PPT)
\cite{peres, horodeccy}.
Note that from the definition (\ref{qft})
it follows that:
\begin{equation}
\widehat{\varrho^T}=\sum_{\i,\j}\varrho_{\i\j}U^{\alpha\beta}_{\i\j}
=\sum_{\i,\j}\varrho_{\i\j}\kappa(U^{\alpha\beta}_{\j\i})^*.
\end{equation}
This suggests the following definition of
a ``transposition map'' $\theta$:
\begin{equation}\label{T}
\theta(a):=\kappa(a)^*.
\end{equation}
Note that, quite surprisingly,
$\theta$ is a homomorphism rather than an
antihomomorphism of the associated Hopf
algebra: $\theta(ab)=\theta(a)\theta(b)$,
but, on the other hand, it is antilinear.
From Propositions \ref{p1} and \ref{p2}
we immediately obtain:

\begin{lem}
Let $\varrho$ act in $\mh_\alpha\otimes\widetilde\mh_\beta$.
Then $\varrho^{T_2}\ge 0$ if and only if 
$(\id\otimes \theta)\hat\varrho$ is positive definite.
\end{lem}

We prove the following analog of the PPT
criterion:

\begin{thm}[Quantum PPT Criterion]\label{ppt}
If an element $a\in\mathcal A\otimes_{\text{alg}}\mathcal B$
is separable, then $(\id\otimes\theta_B)a$,
or equivalently $(\theta_A\otimes\id)a$,
is positive definite.
\end{thm}

First we need the following fact, 
which we prove in Appendix \ref{a}.

\begin{lem}\label{ab} If $a$
is positive definite in $\mathcal A$
and $b$ is positive definite in
$\mathcal B$, then $a\otimes b$ is
positive definite in
$\mathcal A\otimes_{\text{alg}}\mathcal B$.
\end{lem}

Then it is enough to show that if $a$ is positive
definite then $\theta(a)$ is positive definite as well.
Using the notation $\Delta a=\sum_{(a)}a_1\otimes a_2$
one obtains:
\begin{eqnarray}
& &\big(b^*h\kappa\otimes hb\big)\Delta \theta(a)=
\sum_{(a)}\big(b^*h\kappa\otimes hb\big)
\kappa(a_{2})^*\otimes \kappa(a_{1})^*\nonumber\\
& &=\sum_{(a)}h\Big[\kappa\big(\kappa(a_{2})^*\big)b^*\Big]
h\big(b\kappa(a_{1})^*\big)\nonumber\\
& &=\sum_{(a)}\overline{h\Big(\kappa(a_{1})b^*\Big)}\;
\overline{h\Big[b\kappa\big(\kappa(a_{2})^*\big)^*\Big]}\nonumber\\
& &=\sum_{(a)}\overline{h\big(\kappa(a_{1})b^*\big)}\;
\overline{h(ba_{2})}
=\overline{\big(b^*h\kappa\otimes hb\big)\Delta a}\geqslant 0,
\end{eqnarray}
where we used anti-comultiplicativity of $\kappa$ 
(cf. Ref.~\cite{woron_pre}, Proposition 1.9):
$\Delta \kappa=(\kappa\otimes \kappa)\sigma\Delta$
($\sigma$ is the flip), and the
identity: $\kappa\big(\kappa(a)^*\big)^*=a$.$\Box$

Finally, we state the following:
\cite{horodeccy, cmp}:

\begin{thm}[Quantum Horodecki Theorem]
An element $a$ of the quantum group $(A\otimes B, \Delta)$
is separable if and only if 
for every bounded linear map $\Lambda\colon B\to A$ preserving
positive definiteness, $(\id\otimes \Lambda)a$ is positive definite.
\end{thm}

The proof will be given elsewhere. 



\section{A $SU_q(2)$ Example}
Here we present a simple example of 
the transform (\ref{qft}). 
We will consider
a $2\otimes 2$-dimensional quantum system in the
singlet state: 
\begin{equation}\label{psi-}
\Psi_-=\frac{1}{\sqrt{2}}\big(|01\rangle-|10\rangle\big),
\end{equation}
where we use Dirac notation and $|01\rangle$ stands for
the product of the basis elements $e_0\otimes\tilde e_1$, etc.

As the underlying groups
we choose two copies of the 
quantum deformation of $SU(2)$, i.e.
$SU_q(2)$. Recall \cite{woron_pre,woronsuq}
that $SU_q(2)$ is obtained from a universal $*$-algebra
generated by two generators $a,c$ satisfying the relations:
\begin{eqnarray}
& & ac=qca,\ cc^*=c^*c, \ ac^*=qc^*a,\nonumber\\
& &a^*a+\frac{1}{q}c^*c=aa^*+qcc^*=I,
\end{eqnarray}
where $q\in [-1,1]$, $q\ne 0$. Note that for $q=1$ the 
resulting algebra is commutative and one recovers
the standard $SU(2)$ group. Comultiplication
is defined by:
\begin{eqnarray}
& &\Delta(a):=a\otimes a-c^*\otimes c\nonumber,\\
& &\Delta(c):=a\otimes c+c\otimes a^*,
\end{eqnarray}
and coinverse by:
\begin{eqnarray}
& &\kappa(a):=a^*,\ \kappa(c):=-\frac{1}{q}c\nonumber\\
& &\kappa(c^*):=-qc^*, \ \kappa(a^*):=a.
\end{eqnarray}
The fundamental corepresentation 
is given by the unitary matrix:
\begin{equation}\label{suq2}
u=\left[\begin{array}{cc} a & \sqrt{q}c\\
                          -\frac{1}{\sqrt{q}}c^* & a^*\end{array}\right].
\end{equation}
It is enough to consider the fundamental corepresentation 
of $SU_q(2)\times SU_q(2)$. Thus, as the irrep $U$
we take $u\otimes u$. Inserting Eqs. (\ref{psi-}) and
(\ref{suq2}) into Eq. (\ref{qft}) (and stretching the notation a bit)
we obtain:
\begin{eqnarray}
& &\widehat\Psi_-=\langle\Psi_-|u\otimes u \Psi_-\rangle\nonumber\\
& &=\frac{1}{4}\big(a\otimes a^*+a^*\otimes a+c\otimes c^*
+c^*\otimes c\big).
\end{eqnarray}
It is now a highly non-trivial fact, which
follows from our analysis, that the above element
cannot be represented as a convex combination 
of products of positive definite elements.

\section{Conclusions}
This paper follows the research lines of 
Refs. \cite{pra,cmp} and relates the separability problem
in quantum mechanics to abstract problem of 
separability of positive definite functions
on compact groups, and now on compact quantum groups. So far
we have mainly "translated" known results from the 
entanglement theory to harmonic analysis on the 
corresponding groups. In particular, Proposition 3 of 
the present paper point out equivalence of the 
separability of the states and their corresponding 
transforms. We strongly believe that further studies 
of harmonic analysis, in particular in the case of 
finite groups, will allow to obtain novel results 
concerning the separability problem in quantum mechanics. 
One of the goals of this series of papers is indeed 
to stimulate the interest of mathematicians and 
mathematical physicists working in harmonic 
analysis in the separability problem. 

In the course of our analysis we have introduced a notion
of positive definiteness (\ref{posdef2}). There are natural notions of
positive elements in $C^*$-algebras, as well as positive and completely positive
maps of $C^*$-algebras. For ordinary groups, positive definite functions define
positive functionals on the convolution algebra while positive maps correspond
to positive definiteness preserving maps \cite{cmp}. In this context,
note that condition (\ref{posdef2}) can be rewritten as
$((hb)^**hb)a\geqslant 0$, where the convolution of linear functionals
is defined as \cite{woron_pre}
$(\eta'*\eta)a:=(\eta'\otimes\eta)\Delta a$ and the involution as
$\eta^*(a):=\overline{\eta(\kappa(a)^*)}$.

We would like to thank P. M. So\l tan for discussions. 
We acknowledge the support of the EU IP Programme ``SCALA'', 
ESF PESC Programme ``QUDEDIS'', Euroquam Programme FerMix, 
Trup Cualitat programme of the Generalitat de Catalunya,
Spanish MEC grants 
(FIS 2005-04627, FIS2005-04627, FIS2008-00784,
Conslider Ingenio 2010 ``QOIT'', Accionas Integradas). 
J. W. was partially supported by NSF grant
DMS 0623941. M. L. acknowledges Humboldt Foundation.

\appendix\section{Proof of Proposition \ref{ab}}\label{a}
Let $a$, $b$ be positive definite
elements from $\mathcal A$ and $\mathcal B$ respectively.
We use the notation $\Delta a=\sum_{(a)}a_1\otimes a_2$,
$\Delta b=\sum_{(b)}b_1\otimes b_2$
For any finite linear combination
$c=\sum_{i,k}c_{ik}c^A_i\otimes c^B_k$ 
from $A\otimes B$ it then holds:
\begin{eqnarray}
& &\big(c^*h\kappa\otimes hc\big)\Delta(a\otimes b)=\nonumber\\
& &=\sum_{(a),(b)}\big(c^*h\kappa\otimes hc\big)
a_1\otimes b_1\otimes a_2\otimes b_2\nonumber\\
& &=\sum_{(a),(b)}h\big(\kappa_A(a_1)\otimes \kappa_B(b_1)\,c^*\big)
h(c\,a_2\otimes b_2)\nonumber\\
& &\equiv \sum_{i,\dots,l}\overline{c_{ik}}c_{jl}
H^A_{ij}H^B_{kl},\quad\text{where}\label{ajaj}\\
& &H^A_{ik}:=\sum_{(a)}h_A\big(\kappa_A(a_1)c^{A\,*}_i\big)
h_A(c^A_k a_2),
\end{eqnarray}
and analogously for $H^B_{jl}$.
Since $a$ and $b$ are positive definite, $H^A$ and
$H^B$ are positive definite matrices and so is
their tensor product $H^A\otimes H^B$. Hence,
$\sum_{i,\dots,l}\overline{c_{ik}}c_{jl}
H^A_{ij}H^B_{kl}\geqslant 0$. Since any
$c$ from $A\otimes B$ is a norm-limit of
linear combinations of product elements and Haar measure
$h$ is norm continuous, we obtain that
$\big(c^*h\kappa\otimes hc\big)\Delta(a\otimes b)\geqslant 0$
for any $c$.$\Box$


\begin{thebibliography}{99}
\bibitem{mama} R. Horodecki, P. Horodecki, M. Horodecki, and K. Horodecki, arXiv:quant-ph/0702225v2.
\bibitem{schroedinger} E. Schr\"odinger, Naturwissenschaften {\bf 23}, 807 (1935).
\bibitem{bell} J. S. Bell, {\it The Theory of Local Beables}, reprinted in:
J. S. Bell, {\it Speakable and Unspeakable in Qunatum Mechanics}, 
(Cambridge University Press, Cambridge, 2004).
\bibitem{pra} J. K. Korbicz and M. Lewenstein, Phys. Rev. A {\bf 74}, 022318 (2006).
\bibitem{cmp} J. K. Korbicz, J. Wehr, and M. Lewenstein,
Comm. Math. Phys in  press, arXiv:0705.2965v2.
\bibitem{horodeccy} M. Horodecki, P. Horodecki,
and R. Horodecki, Phys. Lett. A {\bf 223}, 1 (1996).
\bibitem{woron_pseudo} S. L. Woronowicz, {\it Compact Quantum Groups},
Warsaw University preprint (1991).
\bibitem{woron_pre} S. L. Woronowicz, Comm. Math. Phys. {\bf 111}, 613 (1987).
\bibitem{piotrek} P. M. So\l tan, S. L. Woronowicz, 	arXiv:0705.2527v2.
\bibitem{u1} Obviously,
transformation (\ref{qft}) can be performed
for any operator from $\mathcal{L}(\mh_\alpha\otimes\widetilde\mh_\beta)$, but
we are interested here only in density matrices.
\bibitem{koornwinder} T. H. Koornwinder, in {\it Representations of Lie groups and quantum groups}, 
V. Baldoni and M. A. Picardello (Eds.), Pitman Research Notes in Mathematics Series {\bf 311} 
(Longman Scientific and Technical, 1994); arXiv:hep-th/9401114v1.
\bibitem{u2} We use a trivial fact:
$\widehat{\varrho_A\otimes\varrho_B}=\hat\varrho_A\otimes\hat\varrho_B$,
following from Eq.~(\ref{prodrep}).
\bibitem{peres} A. Peres, Phys. Rev. Lett. {\bf 77}, 1413 (1996).
\bibitem{woronsuq} S. L. Woronowicz, {\it Twisted SU(2) group. 
An example of a non-commutative differential calculus.},
Publications of RIMS, Kyoto University, {\bf 23}, 117 (1987). 
\end{thebibliography}
\end{document}